# Health-e-Child: An Integrated Biomedical Platform for Grid-Based Paediatric Applications

Joerg FREUND[a] (Project Coordinator), Dorin COMANICIU[a], Yannis IOANNIS[e],
Peiya LIU[a], Richard McCLATCHEY[d], Edwin MORLEY-FLETCHER[b], Xavier
PENNEC[f], Giacomo PONGIGLIONE[c] and Xiang (Sean) ZHOU[a]

On behalf of the Health-e-Child Consortium:
[a]Siemens AG, Erlangen, Germany
[b]Lynkeus SRL, Rome, Italy
[c]IRCCS Giannina Gaslini, Genoa, Italy
University College London, Great Ormond St. Hospital, UK
Assistance Publique Hopitaux de Paris, France
CERN, Geneva, Switzerland
Maat GKnowledge, Toledo, Spain
[d]University of the West of England (UWE), Bristol, UK
[e]University of Athens, Greece
Universita' degli Studi di Genova (DISI), Italy
[f]INRIA, Sophia Antipolis, France
European Genetics Foundation, Bologna, Italy
Aktsiaselts ASPER BIOTECH, Tartu, Estonia
Gerolamo Gaslini Foundation, Genoa, Italy

**Abstract.** There is a compelling demand for the integration and exploitation of heterogeneous biomedical information for improved clinical practice, medical research, and personalised healthcare across the EU. The Health-e-Child project aims at developing an integrated healthcare platform for European Paediatrics, providing seamless integration of traditional and emerging sources of biomedical information. The long-term goal of the project is to provide uninhibited access to universal biomedical knowledge repositories for personalised and preventive healthcare, large-scale information-based biomedical research and training, and informed policy making. The project focus will be on individualized disease prevention, screening, early diagnosis, therapy and follow-up of paediatric heart diseases, inflammatory diseases, and brain tumours. The project will build a Grid-enabled European network of leading clinical centres that will share and annotate biomedical data, validate systems clinically, and diffuse clinical excellence across Europe by setting up new technologies, clinical workflows, and standards. This paper outlines the design approach being adopted in Health-e-Child to enable the delivery of an integrated biomedical information platform.

## 1   Background

From DNA sequencing to laboratory testing and epidemiological analysis, clinicians and researchers produce as well as search for information, as part of their daily routine and decision making. Taking advantage of technology has improved dramatically the quality of these activities' results, facilitating better health-care provision and more advanced biomedical research. Nevertheless, the current state of affairs is still severely restricted with respect to the kind of information that is available to clinicians:

- In each case, clinicians focus their activity around a particular genre of information, e.g., genetic information or laboratory test data, therefore, obtaining a rather narrow and fragmented view of the individual patient that they are examining or the disease that they are investigating.
- For the most part, they are confined to just using information that they themselves or, in the best case, laboratories in their immediate environment generate. For research, they do access general public data banks (e.g., GenBank), but only a limited number of such resources are actually available.
- Especially in paediatrics, longitudinal data (i.e. data taken periodically over time) is usually unavailable and clinicians are forced to operate based on information generated from the current state of their patients.
- Given this fragmented nature of the primary information available, opportunities for large-scale analysis, abstraction, and modelling are very limited as well. Hence, any secondary information and value-added knowledge that comes to the hands of the clinicians is equally restricted.
- Face-to-face conference meetings as well as reading the literature are the only means in the hands of clinicians for exchanging experiences or obtaining second opinions on rare or unclear cases.

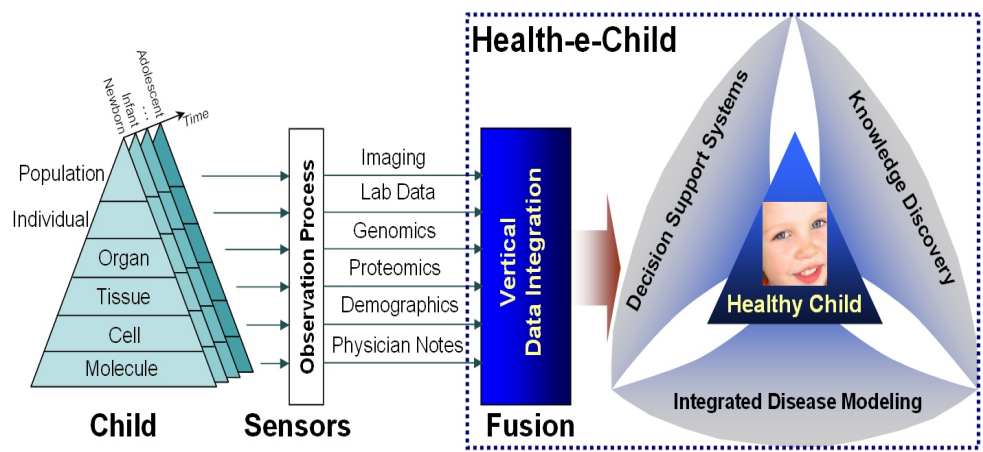

**Figure 1**: Health-e-Child conceptual design

Hence, despite the undisputed advances in biomedical informatics, the above restrictions paint a rather bleak picture of the overall state of affairs. None of the current long-term targets of the field, e.g., personalised medical care, distributed medical teams, multidisciplinary biomedical research, etc. can be realized given the present level of technology support.

The last generation of healthcare projects that have been completed in the EC and in the US have demonstrated the viability of data management techniques based on the Grid [1], [2], [3], [4]. They have shown that it is possible for clinicians to share data and processing between Institutions and even across national boundaries, but they have not addressed the constraints of data heterogeneity, the linkage between biological and medical data, the use of discovered knowledge or handling data that evolves as patients change (especially growth in children). The recently initiated Health-e-Child [5] project is the first step in filling the gap between what is current practice and the needs of modern health provision and research. Its goal is to eventually overcome the above constraints of today's systems and empower clinicians to advance their profession.

Figure 1 illustrates the overall conception of Health-e-Child: today's ever advancing medical sensing technologies generate increasing amounts of information from the children population and encapsulate multiple vertical levels of information from molecular, cell, tissue, to organ, individual, and population level. In particular, the project will develop and test three enabling tool sets for the exploitation of vertically integrated data: disease modelling, decision support systems and services, and knowledge discovery methods and systems. Traditional solutions exist in each of these areas. The novelty of Health-e-Child platform and enabling tool sets lies in the "vertical" aspect:

- The disease models are integrated, i.e., having multiple levels of biomedical information as inputs, including genetic information;
- The decision support systems utilize all biomedical information available for the patient;
- The knowledge discovery modules exploit whatever information is present across multiple heterogeneous databases, including not only traditional but also emerging sources of information, such as molecular or epidemiological data.

## 2  The Health-e-Child Vision

The vision is for the Health-e-Child system to become the universal biomedical knowledge repository and communication conduit for the future, a common vehicle by which all clinicians will access, analyze, evaluate, enhance, and exchange biomedical information of all forms.

Clearly, any effort towards this vision requires significant change in the biomedical information management strategies of the past, with respect to functionality, operational environment, and other aspects. Contrary to current practice, the vision requires that the Health-e-Child system be characterized by the following:

1. Universality of information: Health-e-Child should handle "all" relevant medical applications, managing "all" forms of biomedical content. Such breadth should be realized for all dimensions of content type: biomedical abstraction (from genetics, to clinical, to epidemiological), temporal correlation (from current to longitudinal), location origin (any hospital/clinical facility), conceptual abstraction (from data to information to knowledge), and syntactic format (any data storage system, from structured database systems to free-text medical notes to images).
2. Person-centricity of information: Health-e-Child should synthesize all information that is available about each person in a cohesive whole. This should form the basis for personalised treatment of the individual, for comparisons among different individuals, and for identifying different classes of individuals based on their biomedical information profile.
3. Universality of application: Health-e-Child should comprehensively capture "all" aspects of "all" biomedical phenomena, diseases, and human clinical behaviours. This includes growth patterns of healthy or infected organic bodies, correlations of genotype/phenotype under several conditions of health, normal and abnormal evolution of human organs, and others.
4. Multiplicity and variety of biomedical analytics: Health-e-Child should provide a rich and broad collection of sophisticated analysis and modelling techniques to address the great variety of specialized needs of its applications. It should synthesize several suites of disease models, decision trees and rule systems, knowledge discovery and data mining algorithms, biomedical similarity measures, ontology integration mappings, and other analytical tools so that clinicians may obtain multi-perspective views of the problems of concern.
5. Person-centricity of interaction: The primary concern of any user interaction with Health-e-Child should be the persons involved. This should be realized at three levels at least. First, the system should facilitate clinicians in identifying or generating easily all information that is pertinent to their activity and should only offer to them support for their decision making and not direct decisions. Second, it should protect the privacy of the person whose data is being accessed and manipulated. Third, it should allow biomedical information exchanges and information-based collaborations among clinicians.
6. Globalness of distributed environment: Health-e-Child should be a widely distributed system, through which biomedical information sources across the world get interconnected to exchange and integrate their contents.
7. Genericity of technology: For economy of scale, reusability, extensibility, and maintainability, Health-e-Child should be developed on top of standard, generic infrastructures that provide all common data and computation management services required. In the same spirit, all integration, search, modelling, and analysis functionality Health-e-Child incorporates itself should be based on generic methods as much as possible. Any specialized functionality should be developed in a customized fashion on top of them.

The Health-e-Child proposal aims at developing a first version of the vision. It focuses on key instances of the above, the emphasis of the Health-e-Child effort being on "universality of information" and its corner stone is *the integration of information across biomedical abstraction, whereby all layers of biomedical information (i.e., genetic, cell, tissue, organ, individual, and population layer) are vertically integrated to provide a unified view of a person's biomedical and clinical condition.*

This drives the research and technology directions pursued with respect to all other target features of Health-e-Child, in particular, temporal and spatial information integration, information search and optimization, disease modelling, decision support, and knowledge discovery and data mining, all operating in a distributed (Grid-based) environment. Each one of these areas presents novel technical challenges, which the current state of the art cannot meet, when the relevant issues are examined in conjunction with vertically integrated biomedical information.

## 3 Project Aims and Objectives

The general objectives of Health-e-Child are the following:
- To gain a comprehensive view of a child's health by vertically integrating biomedical data, information, and knowledge, that spans the entire spectrum from genetic to clinical to epidemiological;
- To develop a biomedical information platform, supported by sophisticated and robust search, optimization, and matching techniques for heterogeneous information, empowered by the Grid;
- To build enabling tools and services on top of the Health-e-Child platform, that will lead to innovative and better healthcare solutions in Europe:
    a. Integrated disease models exploiting all available information levels;
    b. Database-guided biomedical decision support systems provisioning novel clinical practices and personalised healthcare for children;
    c. Large-scale, cross-modality, and longitudinal information fusion and data mining for biomedical knowledge discovery.

The project focus will be on individualized disease prevention, screening, early diagnosis, therapy and follow-up of paediatric heart diseases, inflammatory diseases, and brain tumours. The project will build a Grid-enabled European network of leading clinical centres that will share and annotate biomedical data, validate systems clinically, and diffuse clinical excellence across Europe by setting up new technologies, clinical workflows, and standards. Paediatrics adds a temporal dimension along which biomedical information changes at different speeds for the different layers of biomedical abstractions, whose vertical integration therefore faces further challenges. The particular diseases in the categories chosen correspond to largely uncharted territories with significant impact expected by any major advances in our understanding of them; they also represent a broad spectrum of technology requirements, thus ensuring genericity and broad applicability of the end result.

## 4 Data Integration Philosophy

One important aim of the Health-e-Child project is to provide an integrated healthcare platform for European paediatrics and this paper outlines first ideas in how this platform will be designed. As stated earlier, this platform will enable the modelling and integration of relevant biomedical sources across different diseases or patient levels and the development of a Grid-based service-oriented environment to manage distributed and shared heterogeneous biomedical data and knowledge sources. It will also enable the use of integrated decision support and knowledge discovery systems but this is beyond the scope of this architecture paper.

The main data management effort in Health-e-Child is focused on building a comprehensive data, medical information and knowledge-discovery infrastructure for various higher-level components of the Health-e-Child system as recommended in [6], [7]. The design philosophy is founded on four cornerstones:
1. the Grid middleware, which provides the virtual foundation for flexible, secure, coordinated sharing of distributed resources.
2. the modelling and integration of relevant biomedical data sources for improved medical knowledge discovery and understanding

3. a Grid-enabled service-oriented gateway that is responsible for data access and management of Health-e-Child acquired and integrated data and
4. a medical query processing environment to provide necessary indexing, search and processing facilities, in the form of algorithms, methods and metrics, for identifying information, knowledge and data fragments that are relevant to a particular request.

The biomedical data sources referred to in point 2 cover several vertical levels (from cellular information through organ information to patient and population information) and Health-e-Child will develop data and knowledge models integrating across these levels. This core project effort we refer to as "vertical data integration in the medical domain" and it mainly covers point 2 above.

Ontologies [8], [9] will be used to formally express the Health-e-Child medical domain, for improved communication of domain concepts among domain components, and to assist in the integration process. Moreover, ontologies provide semantic coherence of the integrated data model, as ontological commitments will be expected from the Health-e-Child components. In the first few months of the project ontology software and other software technologies are being acquired, tested and evaluated in the context of the project's user requirements, especially with regard to the integration of clinical data. The ontology-guided semantic integration for generating case data will investigate the following research stages: mapping discovery, the declarative formal representation of and reasoning with mappings.

**4.1 Mapping Discovery**

This stage covers identifying similarities between Ontologies in order to determine which concepts and properties represent similar notions across heterogeneous data samples (semi-) automatically. One of major bottlenecks in generating viable integrated case data is that of mapping discovery. There exist two major approaches to mapping discovery:
1. A Top-down Approach. This approach is applicable to ontologies with a well-defined goal. Ontologies usually contain a generally agreeable upper-level (top) ontology by developers of different applications; these developers can extend the upper-level ontology with application-specific terms. Examples from this approach are Suggested Upper Merged Ontology (SUMO) [10] from the IEEE Standard Upper Ontology Working Group and DOLICE [11].
2. A Heuristics Approach. This approach uses lexical structural components of definitions to find correspondences with heuristics. For example, [12] describes a set of heuristics used for semi-automatic alignment of domain ontologies to a large central ontology. PROMPT [13] supports ontology merging, guides users through the process and suggests which classes and properties can be merged. FCA-Merge [14] supports a method for comparing ontologies that have a set of shared instances. IF-Map [15] identifies the mappings automatically by the information flow and generates a logic isomorphism [16].

Based on medical ontologies e.g. [17], [18], [19], Health-e-Child will investigate the mapping heuristics for integrated case data. We will evaluate the relative quality of several of these mapping discovery methods for integrated case data. Then we will provide an optimal combination of the best methods with respect to the accuracy and computation-times.

## 4.2 Declarative Representation of Mappings

This element represents the mapping between ontologies in order to enable reasoning with mappings. The higher expressive power of medical ontology representation language provides opportunities to represent the mappings in more expressed terms. There are several approaches in mapping representations: instance-based [15] and [20], axiom-based [21] and view-based [22]. In axiom-based mapping representation, the correspondence between two ontologies is expressed as a set of formal logical axioms relating to classes and properties of two source ontologies. Thus, logical axioms are essentially deduction rules for bridging two ontologies by relating concepts from one source ontology to another ontology, by using a theorem prover for mapping deduction. In instance-based mapping representation, the correspondences between two ontologies are declaratively represented as transformation functions of instances. Thus, mappings are essentially transformation rules for linking source ontologies to targets. In view-based mapping representation, the correspondences are represented as views, similar to database views definitions in information integration. Thus, the correspondences between two ontologies are defined as a query in terms of views definitions. We will investigate the logic-approach to provide combination of mapping representations methods in a general logic framework.

## 4.3 Reasoning with Mappings

The final stage considers actions on the mappings once they are defined and the identification of types of reasoning that can be developed to generate integrated case data. Many semantic integration tasks based on reasoning have been proposed: ontology projection and extension [21], ontology merging [13] and Description Logics (DL) reasoning [22]. We will investigate the ontology linkage to support the generation of suitable integrated case data.

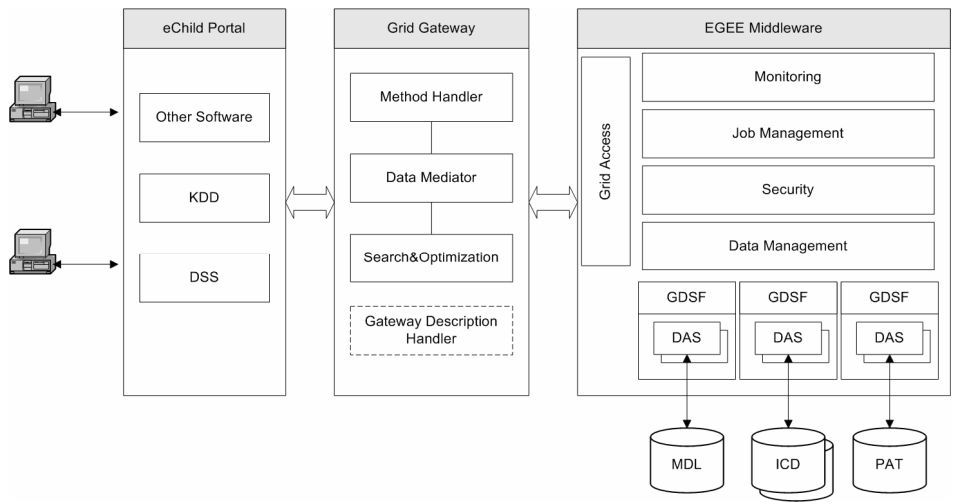

**Figure 2:** Initial Health-e-Child Data Architecture

## 5    Initial Architecture Design

Figure 2 shows a high level representation of the draft Health-e-Child architecture. The Grid Gateway, the EGEE Grid Middleware and the Grid Data Service Factory (GDSF) represent the main software components of data management. The high-level decision support and knowledge discovery components (DSS and KDD of figure 1) are served by the Health-e-Child Portal, with the Grid Gateway providing the infrastructure for the use services for accessing and sharing of the distributed data and knowledge on the Grid. Data includes Application Models (MDL), Integrated Case Data (ICD), Pattern and Trend data (PAT) retrieved through a Data Access System (DAS).

Currently the EGEE software [23] is envisaged to provide the Grid platform since it will provide 'standardised' access to components for Grid Access, Monitoring, Job Management, Security and Data Management. Health-e-Child will be designed to provide services which interface to generic Grid services thereby allowing use of EGEE when it becomes available and providing a migration path to future Grids standards as and when they appear, a philosophy that was previously shown to be a success in the MammoGrid project [3]. This would also enable the interoperation of the Health-e-Child application with other Grids middleware as in [6].

Health-e-Child aims to deliver a set of semantically rich models for integrated biomedical data and knowledge. The models produced will be novel in that they will unify semantically remote layers (from genetic through to clinical) thus spanning a domain which has never before been fully captured. This will rely heavily on the expertise of data modellers, computer science experts in data and knowledge modelling, domain experts and clinical partners and will take advantage of the advances made by the INFOGENMED [24] project as described in section 6. It will provide a set of analysis and design models that facilitates the integration of relevant biomedical sources for improved medical knowledge discovery and understanding.

The integration and modelling activities of Health-e-Child will take into consideration issues that are typically addressed by data integration projects (in particular we shall also investigate the role of 'wrappers' and/or 'translators' in the semantic integration of biomedical information). Foremost is the issue of heterogeneity with respect to the source data schemata and database management system products hosting the database. The Health-e-Child project will have several medical institutions contributing diverse biomedical data for the different vertical levels. It is likely that data sources for each level will have different schemata, using different software packages and varying types of access controls. A necessary first step towards bringing these disparate sources together is to identify the core entities for each level, and an intermediary data model per level will be proposed to capture the entities' structures. This effort will rely on reviewing and adopting existing ontologies of the subdomains and will benefit from work in other current biomedical projects such as MYGRID [25] and in initiatives such as WSMF [26]. The next step is to identify the relevant structures to unify these six levels (i.e. genetic, cell, tissue, organ, individual, and population) into a single data model whose semantics is captured by the integrated ontology. The integrated data model captures the structural representation of the level entities and their relationships, and provides a coherent view of the integrated domain.

In addition to heterogeneity, other issues related to integrating distributed heterogeneous data sources will be addressed. Some of these issues are data-related (e.g. distribution, acquisition, normalisation, aggregation, curation), access-related (e.g.

transaction management, query processing), network-related (e.g. location, fragmentation), and privacy and security. Existing integration solutions will be consulted including data mediators, layered and meta-modelling approaches (including the CRISTAL [27] approach used in the MammoGrid project). Medical standards (e.g. HL7, DICOM, H12) and informatics standards (e.g. UML, XML, SQL, ODMG) will be followed as closely as possible, where applicable and feasible.

To ensure the semantic coherence of the integrated data model, the formal concepts will be documented as ontological commitments. The ontological foundation of the analysis model will ensure the consistency of the design model used for the data integration and will form the basis for the medical knowledge management system. The formal conceptualisation of semantics will rely on existing medical ontologies and the relationship to those will be documented. The emerging framework of semantic models will be the foundation for the knowledge model behind the knowledge sharing and mining infrastructure of the Health-e-Child system.

The Grid Gateway is responsible for data management and distribution; it provides access to data mining, knowledge discovery and optimisation algorithms; it incorporates and integrates raw medical data from various data sources. For client user interfaces and decision support components, the API and services of the Health-e-Child Grid Gateway provide an abstraction of the underlying Grid middleware, data resources and data management mechanisms. The data management layer will comprise a set of services for exploiting the information supplied by different medical information repositories. These services offer the following functionality:

- Information and knowledge extraction. The information includes amongst others: metadata that can constitute part of the external schema, temporal attributes used for content tracking, as well as keywords to be used for indexing and content querying;
- Repository content tracking. This component registers both the changes discovered in the stored data items, and the modifications of the schemata of the data sources. The tracking mechanism is used to maintain the integrity of the information and the external global schema;
- A security level processor. This is in charge of ensuring user privileges coherence and provides mechanisms to define user profiles over the global schema. The restrictions imposed on the global schema are translated later into local data source restrictions and vice versa. The security mechanism will extend the one provided by the underlying Grid middleware to fit the particularly strict requirements of the medical domain;
- Integration of applications and different databases like Oracle, DB2, LDAP directories, etc.;
- Global database management: integrity functions, corruption detection, index rebuilding, etc.;
- A data view which is fully conformant to the model incorporating vertical medical integration developed by WP6.

## 6   Related Projects

Initiatives from which Health-e-Child is expected to benefit include the BIRN [28] project in the US, which is enabling large-scale collaborations in biomedical science by utilizing the capabilities of emerging Grid technologies. BIRN provides federated medical data, which enables a software 'fabric' for seamless and secure federation of data across the network and facilitates the collaborative use of domain tools and flexible processing/analysis frameworks for the study of Alzheimer's disease. The INFOGENMED initiative [24] has given the lead to projects in moving from genomic information to individualized healthcare and Health-e-Child will build on its findings in vertical data modelling. The IHBIS project [2] has proposed a broker for the integration of heterogeneous information sources in order to collect protect and assemble information from electronic records held across distributed healthcare agencies. This philosophy is one that will also be investigated in Health-e-Child. In addition, the fame-permis project [29] is currently in the process of developing a flexible, authentication and authorization framework to cope with security issues for a healthcare environment; aspects that are important in the delivery of the Health-e-Child prototypes. Finally the CDSS [1] project being a system that uses knowledge extracted from clinical practice to provide a classification of patients' illnesses, implemented on a Grid clearly impacts the decision support elements of the Health-e-Child project.

Furthermore, the MYGRID project [25] is one which indicates the benefits of an ontological approach to federated data access on the Grid in the bioscience domain. MYGRID uses a web services approach as its underlying distributed systems infrastructure with an intention to migrate to Globus/OGSA based solutions at a later date. They use OWL [30] for the ontology language using description logic as opposed to the emerging WSMO (Web Service Modelling Ontology), based on first order logic, proposed recently [31]. WSMO is based on the Web Service Modelling Framework [32] and will enable the realization of true semantic web services, the next step in allowing Grid-based ontology mediation. Such developments are the first step in the provision of autonomous Semantic Grid systems. By adopting an ontology-based solution to unifying genetic/genomic data to patient/clinical data, it is expected that the Health-e-Child project will take an active role in influencing the future of biomedical ontology-based Grid solutions.

## 7   Conclusions – The Way Ahead

To reach its goals, Health-e-Child must innovate in diverse scientific areas. In particular, the scientific and technologic objectives of Health-e-Child are to advance the state-of-the-art in the following areas:

- Translation, mapping, and matching of biomedical metadata (ontologies and other semantic metadata forms as well as syntactic structures) for vertical data integration;
- Vertically integrated information modelling and knowledge representation;
- Personalised models integrating all information sources about a patient's health and disease biases;
- Multi-objective optimization of biomedical information search;
- Efficient similarity indexing and information retrieval from vertically integrated biomedical databases;
- Autonomous, consistent, un-biased, and validated biomedical data analysis algorithms;

- Cluster analysis, feature sensitivity analysis, data mining and association, and knowledge discovery from vertically integrated biomedical data;
- Robust information fusion algorithms leading to new decision methods from integrated biomedical data;
- Personalised methods for risk assessment, diagnosis, prevention and therapy; robust statistics and multiple hypothesis prediction for assessment of therapeutic response.

The novel techniques resulting from the research in the areas above will be implemented and incorporated into the overall Health-e-Child system. The latter will be a distributed data and computation management system based on the Grid architecture. It will be built on top of the middleware developed as part of the EGEE project and, for security reasons, it will operate in an infrastructure that will be private to the project. Primary information will be collected at the sites of the three Hospital partners and, after appropriate anonymization and other necessary forms of curation, will be available for manipulation and processing by the rich stack of Health-e-Child modules. At the other end, clinicians will drive the system, using enabling tools to obtain second opinions on particular clinical cases or to identify interesting patterns in the data available while studying particular phenomena as part of biomedical research. In both scenarios, Health-e-Child will be a powerful tool in the hands of the clinician, bridging the gap between the latter's conception of the biomedical problem at hand and the information available supporting the various alternatives for its solution.

This paper has outlined the challenges facing the Health-e-Child project and has identified the project aims and objectives and highlighted its design strategy. In addition it has indicated the first steps being taken in the project to deliver an integrated platform for paediatrics that will become the foundation for future Grid-based biomedical solutions.

**Acknowledgements**

The authors thank the European Commission and their institutes for support and particularly acknowledge the contribution to this paper of the following Health-e-Child consortium members: Alok Gupta (Siemens AG), Alberto Martini and Paolo Toma (IRCCS Giannina Gaslini, Genoa), Younes Boudjemline (Assistance Publique Hopitaux de Paris), Catherine Owens (Great Ormond St Hospital, London), Florida Estrella (CERN), David Manset and Alfonso Rios (Maat GKnowledge), Tamas Hauer and Dmitry Rogulin (UWE), Alessandro Verri (DISI) and Alessandro Sattanino (Lynkeus).